\documentclass[runningheads]{llncs}
\usepackage{epsfig}
\usepackage{amsmath}
\usepackage{amssymb}
\usepackage{subfig}
\usepackage{multirow}
\usepackage{boldline}

\begin{document}

\title{CD$^2$: Combined Distances of Contrast Distributions for Image Quality Analysis}
%
%
\titlerunning{CD$^2$: Combined Distances of Contrast Distributions for IQA}
\author{Sascha Xu\inst{1} \and
Jan Bauer\inst{2} \and
Benjamin Axmann\inst{3}\and
Wolfgang Maass\inst{4}}
\authorrunning{Xu et al.}
%
\institute{X-Motive, Germany \email{sascha.xu@xmotive.de}
\and
Hochschule Karlsruhe, Germany
\email{jan.bauer@hs-karlsruhe.de}\\
\and 
Mercedes-Benz, Germany
\email{benjamin.axmann@daimler.com}\\
\and 
DFKI, Germany
\email{wolfgang.maass@dfki.de}\\
}

\maketitle              
\begin{abstract}
The quality of visual input impacts both human and machine perception.
Consequently many processing techniques exist that deal with different distortions. 
Usually they are applied freely and unsupervised. 
We propose a novel method called CD$^2$ to protect against errors that arise during image processing. 
It is based on distributions of image contrast and custom distance functions which capture the effect of noise, compression, etc.
CD$^2$ achieves excellent performance on image quality analysis benchmarks and in a separate user test with only a small data and computation overhead.

\end{abstract}
\section{Introduction}
\begin{figure}[b]
\centering
\subfloat[Street with Pedestrians and Cars]{
\centering
\includegraphics[width=0.45\textwidth]{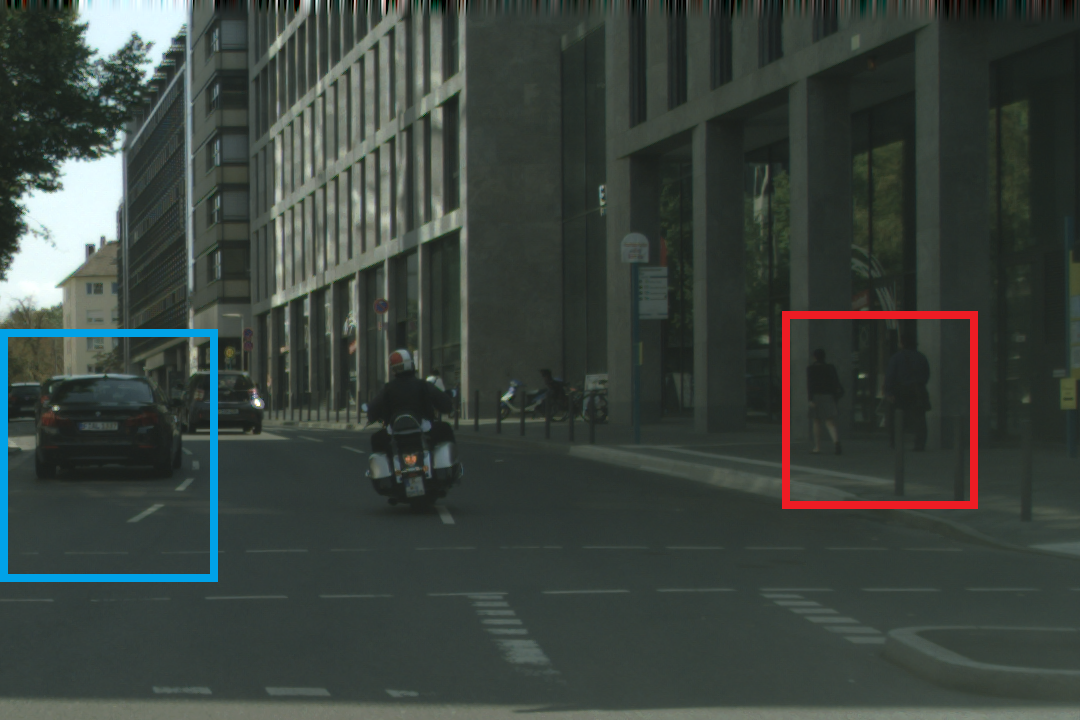}
}
\subfloat[Processed Street Scene]{
\centering
\includegraphics[width=0.45\textwidth]{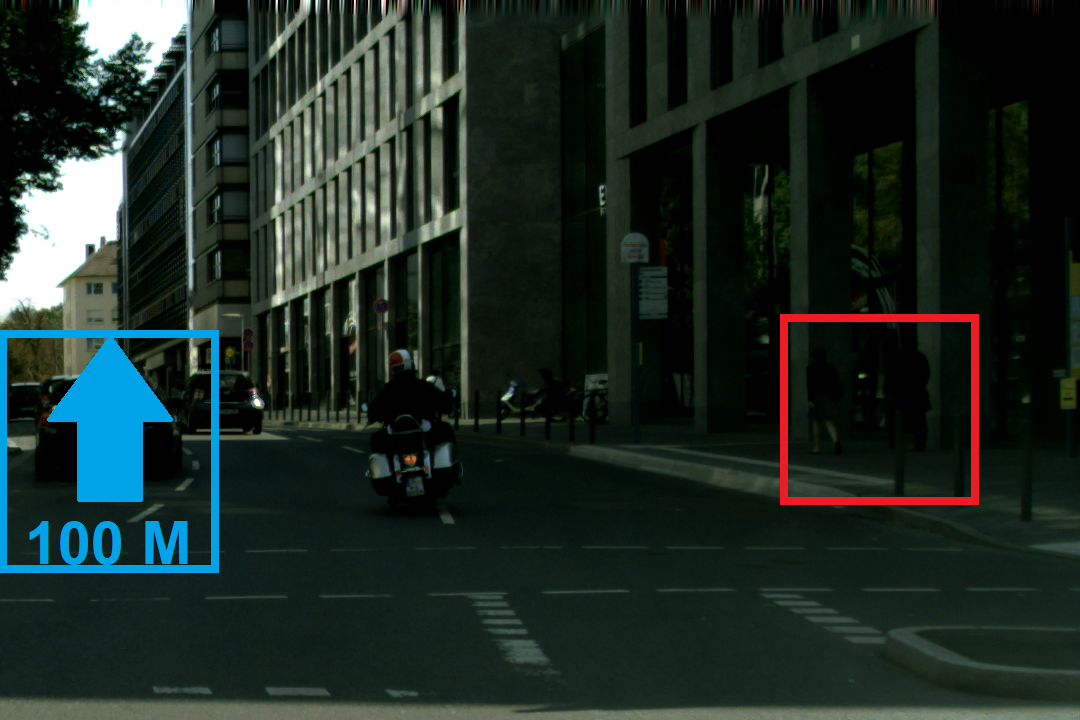}
}

\caption{Safety Hazard: Misused Image Processing}
\label{fig:proc_failures}
\end{figure}
Modern cars have up to 8 cameras giving full coverage of the traffic situation. 
Their video data is relevant to both human drivers and autonomous driving agents.
In accordance with automotive functional safety guidelines \cite{iso26262} we propose a novel method for guarded image transmission and processing.
The perceptual differences between an original image and a processed version have to be assessed.
The goal is to detect safety hazards that hinder the extraction of information about items in the scene.
Image processing explicitly changes pixel values thus methods such as the CRC-checksum are not applicable.
Instead a perceptual comparison has to be conducted for the image pair:
Where and what kind of changes have been applied to the image?
Have objects been obscured or even lost?
Is the image still perceptually similar to the reference?

Safety hazards shall be detected so that warning and countermeasures can be taken.
A manifestation of safety hazards in image processing is displayed in figure \ref{fig:proc_failures}.
Because of a contrast enhancement operation, the pedestrians in the red box have become poorly visible.
In the blue box image augmentation covers up a vehicle and could cause the driver to overlook it.
In general wrong processing in wrong moments may lead to accidents by hindering/misleading the driver.

Algorithms are affected by this as well: 
Dodge and Karam \cite{dodge2017study} examine the effect of noise, blur and compression distortions on neural networks.
They show that widely used networks perform much worse on the Imagenet \cite{deng2009imagenet} challenge when faced with distorted images. 
Thus on the one hand there is the need to process blurry and noisy camera data for driver assistance systems.
On the other hand it must be ensured that the image processing itself does not become a source of error.

\subsection{Related Work}
This task fits very well into the domain of image quality assessment (IQA).
There exist three modi operandi for IQA: Full-reference IQA (FR-IQA) has access to the complete reference and result image.
The structural similarity index (SSIM) by Wang et al.~\cite{wang2004ssim} is the most prominent FR-IQA index . 
SSIM assesses image similarity with pixel statistics about luminosity, contrast and structure.
Since visual perception is illumination invariant and instead relies on luminance differences and patterns, subsequent methods such as the Gradient-SSIM of Chen et al.~\cite{chen2006gradient_ssim} operate on gradient maps.
Sampat et al.~\cite{sampat2009wave_ssim} utilize the Wavelet transformation to represent structures of an image and create their variant Wavelet-SSIM.
Finally the latest FR-IQA indices such as FSIM by Zhang et al.~\cite{zhang2011fsim} are built on a combination of Wavelet and gradient similarity measures which are pooled together into one quality index.

Blind IQA aims to judge the quality of an individual image. 
There is no reference image, only the result image is available. 
Traditionally this was approached similar to full-reference methods using features from the frequency and gradient domain  \cite{saad2012blind,xue2014blind} . 
With the rise of neural networks CNN architectures have been utilized for blind IQA as well \cite{kang2014convolutional}.
For the evaluation of image augmentation however blind IQA is not suitable.
To make a qualified statement about what parts of the image are covered up, the content of the reference image must be taken into account.

The middle ground between full-reference and blind IQA is occupied by reduced-reference IQA (RR-IQA). 
Here the use of natural scene statistics and distributions to represent images is widespread \cite{li2009reduceddivisenormal,wang2005reducedwavstat,tao2009reducedcont}. 
By applying statistical distances on the distributions of reference and processed image RR-IQA requires only limited data.

The goal of this paper is to develop a method to safeguard and control results of image processing and transmission. 
Therefore the reference image should be taken into account. 
A full-reference method is costly in bandwidth and often not feasible for implementation. 
Consequently we opt for a reduced reference approach.

\begin{figure}[b]
    \centering
    \includegraphics[width=\textwidth]{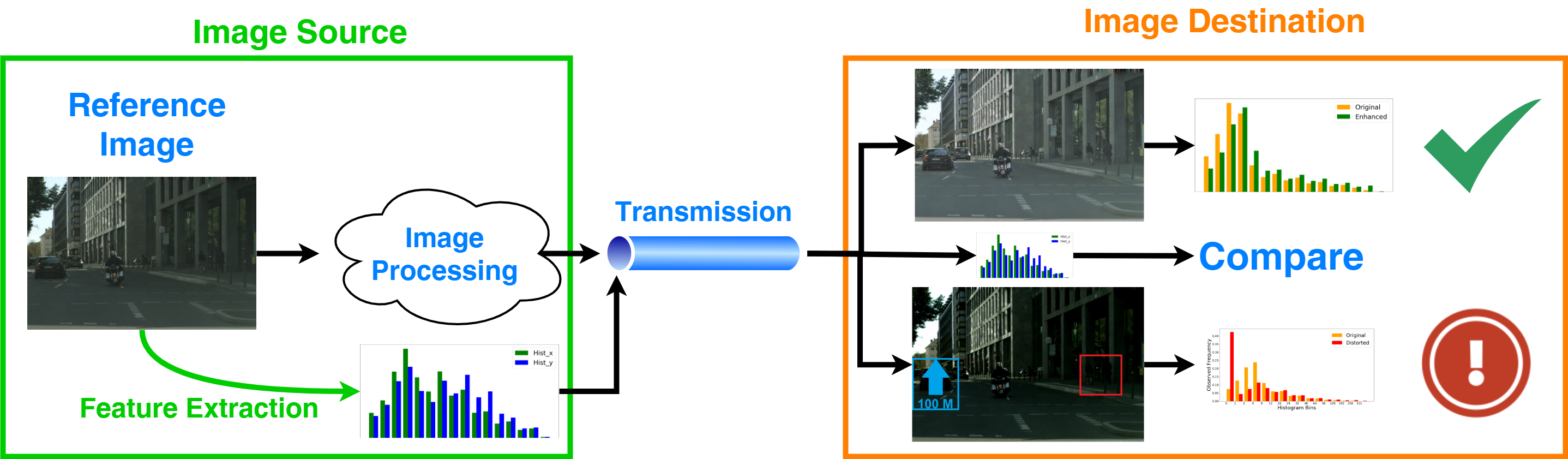}
\caption{Layout of a Guarded Image Processing System}
    \label{fig:flow_diagram}
\end{figure}
\subsection{Guarded Image Processing}
The layout of the proposed method is displayed in figure \ref{fig:flow_diagram}.
An image source (e.g.\ rear-view camera) produces the reference image which is to be processed and transmitted to its destination, typically a display or a driver assistance system.
The reference image is represented through a feature set.
The feature set acts as a signature for the image, which should contain sufficient information but in much reduced size.
Motivated by the affinity of the human visual system for contrast we utilize histograms to capture the distribution of contrast in local regions of the image.
The features are sent along the image stream and travel through the processing pipeline to the destination.

At the destination the feature set of the processed image is compared to the features of the reference.
\textbf{CD$^2$} stands for the combined use of distance functions that measure general and distortion specific similarity of the contrast distributions.
In the end a detailed map for local distortion as well as an accurate global indicator is obtained.
This allows to diagnose and locate dangerous image processing safety hazards and create a strong, resilient system.
CD$^2$ offers the following advantages compared to current RR-IQA methods:
\begin{enumerate}
    \item \emph{Improved Perceptual Modelling}: RR-IQA methods rely on statistics of known factors like gradients and wavelets. We further explore the connection of their distributions to perception by incorporating them into a hierarchical model for the different stages of visual perception.
    \item \emph{Local Predictions}: Telltales affect only a small region of the image, but IQA methods are designed for entire images. We design our method to make precise local comparisons.
    \item \emph{Efficiency}: In real world application hardware requirements should be low. The FPGA prototype of CD$^2$ is able to operate on pixel streams without frame buffering and allows real time inference at 60fps.
\end{enumerate}

Evaluation is carried out through a user test on automotive displays focusing on distortion of street scenes.
It is complemented by benchmark tests on the LIVE\cite{sheikh2005live}, TID \cite{ponomarenko2015tid} and CSIQ \cite{larson2010csiq} IQA databases.
In the end we find that the proposed method performs better than competing RR-IQA methods at a very low time and hardware complexity.
The rest of the paper is organized as follows: Section 2 introduces our features. 
Section 3 focuses on the distance functions used to compare to images. 
Section 4 details the experiments and the user test. 
Section 5 describes the FPGA prototype and gives closing
remarks. 

\section{Contrast Distribution as Features}
To ensure safety of image processing we impose the following requirement:
Objects which were visible in the reference image must be visible in the altered image.
Therefore the main focus lays on the early stages of visual perception when the shapes and contours of objects are detected.
Tschechne and Neumann \cite{tschechne2014hierarchical} propose a hierarchical approach to model the process of shape recognition in the visual cortex.
We employ a simplified feedforward model which models the factors that allow our visual system to detect objects, but not the specific shapes themselves.
Our three stage model tries to account for pixel scale contrast stimuli, intermediate stage grouping of stimuli and a global weighting of regions into a complete picture.

Image contrast as a visual property is closely associated to stimulus strength for our visual system.
During the detection of objects these contrasts in the form of edges and shadows play a major role \cite{avidan2002contrast}.
Algorithmic edge detection commonly employs image gradients, which are also a great indicator of blur.
Therefore they serve as the contrast measure of our choice.
The gradients will be computed on the luminance channel of the CIELAB color space.
For an image $I(x,y)$ consisting of luminance integers between 0 and 255, the image gradient $\nabla I(x,y)$ may be approximated with a number of kernels.
In the following 3x3 Sobel kernels are used to filter an image.
The resulting pixelmap of contrast stimuli/gradients usually amounts to millions of values per image.

\begin{equation}
    \nabla I(x,y) = [\frac{\delta I(x,y)}{\delta x},\frac{\delta I(x,y)}{\delta y}] = [g_x(x,y),g_y(x,y)]
\end{equation}

\begin{figure}[t]
\centering
\subfloat[Division of Image into Patches]{
\includegraphics[width=.7\textwidth]{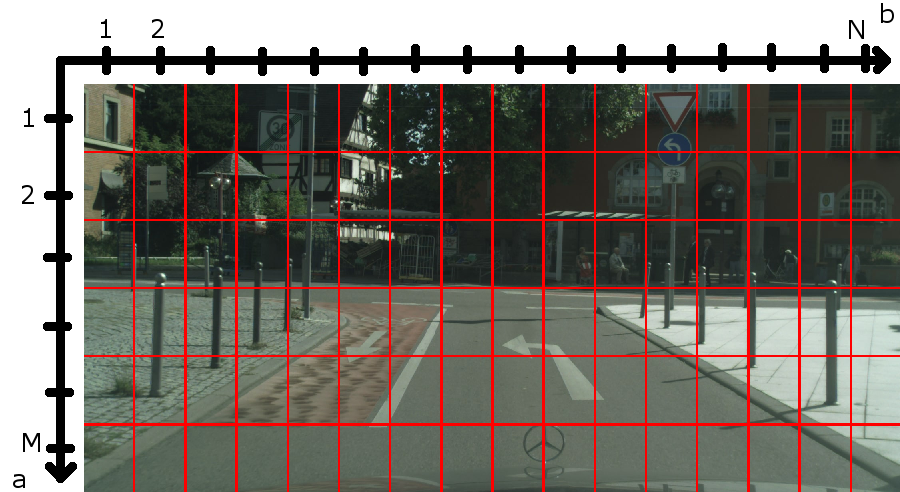}
\label{fig:patches}
}

 \centering 
 \subfloat[Distribution of Gradients]{
 \includegraphics[width=.5\textwidth]{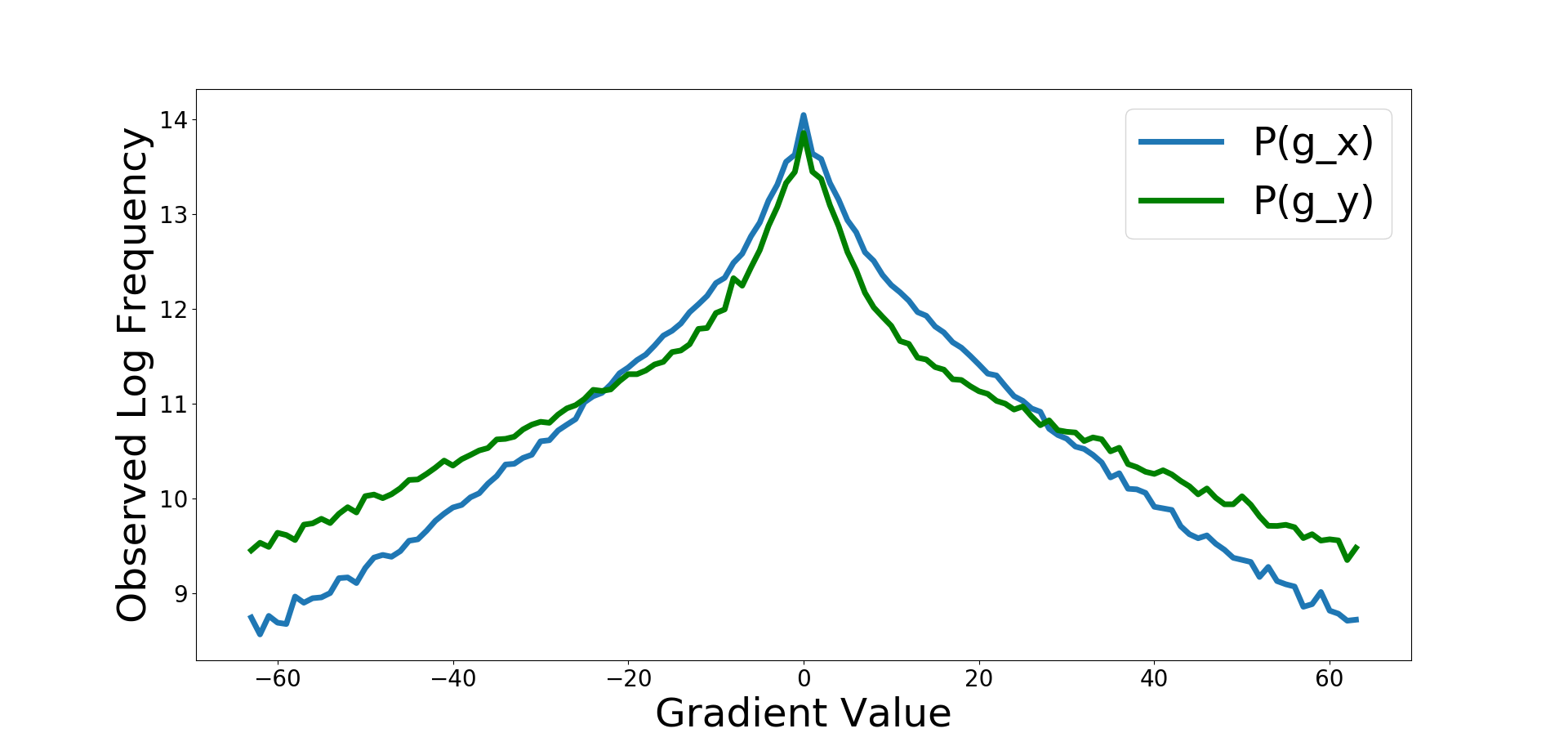}
  \label{fig:grad_dist}
 }
 \centering 
 \subfloat[Distribution of Absolute Gradients]{
 \includegraphics[width=.5\textwidth]{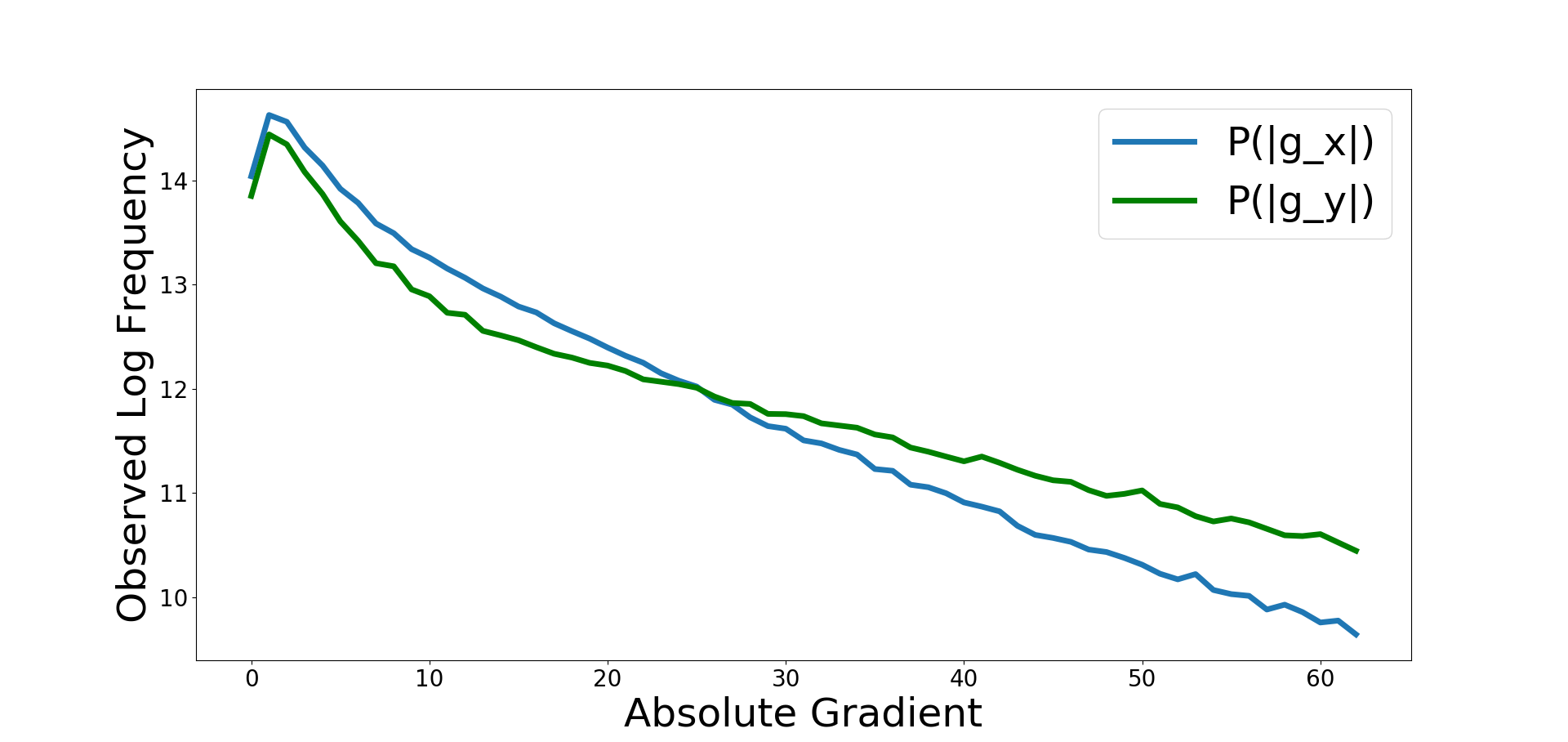}
 \label{fig:abs_dist}
 }
 \caption{Empirical Gradient Distributions for Reference Image}
 \label{fig:feature_figure}
\end{figure}

Objects span over larger areas and thus many, many pixels.
The visual system scans over areas and searches for evidence of objects which manifest themselves through edges and shadows.
One local gradient value alone may not mean much unless it is put into context with surrounding gradients.
Therefore the image is divided up into a evenly spaced grid of MxN patches shown in figure \ref{fig:patches}.
Each image patch in row a, column b is analyzed by its gradient distributions $P_{ab}(g_x)$ and $P_{ab}(g_y)$.
They contain vital information regarding edges, surfaces, shadows and allow to detect distortions like noise or compression through comparisons (topic of section 3).
The sign of the gradients is dropped based on the assumption that natural image gradients are zero symmetric distributed (see  figure \ref{fig:grad_dist} and \cite{huang1999statistics}).
This natural scene statistics approach condenses the gradients into a much more compact form while retaining sufficient information to make a qualitative prediction about differences and deterioration of image quality.

\subsection{Gradient Distributions to Contrast Histograms}
The gradient distribution is defined for all possible unsigned Sobel gradients.
This makes it hard to process and unnecessary large in size.
Histograms help categorize gradients into similar segments and shrink down the domain of the distribution.
Here the challenge lies in dividing up the domain into segments of similarly perceived stimuli.
Though we deal with gamma encoded pixel values, a linear division of the domain does not seem appropriate.
As seen in figure \ref{fig:grad_dist} the gradients have a center heavy distribution.
Huang et al. \cite{huang1999statistics} explain this phenomenon with the prevalence of homogeneous surfaces and sparsity of edges in natural scenes, resulting in exponentially decreasing probability for linearly increasing gradient.

\begin{figure}[t]
{
\centering
\subfloat[$Hist_{11}(g_x)$ and $Hist_{11}(g_y)$]{
\includegraphics[width=.5\textwidth]{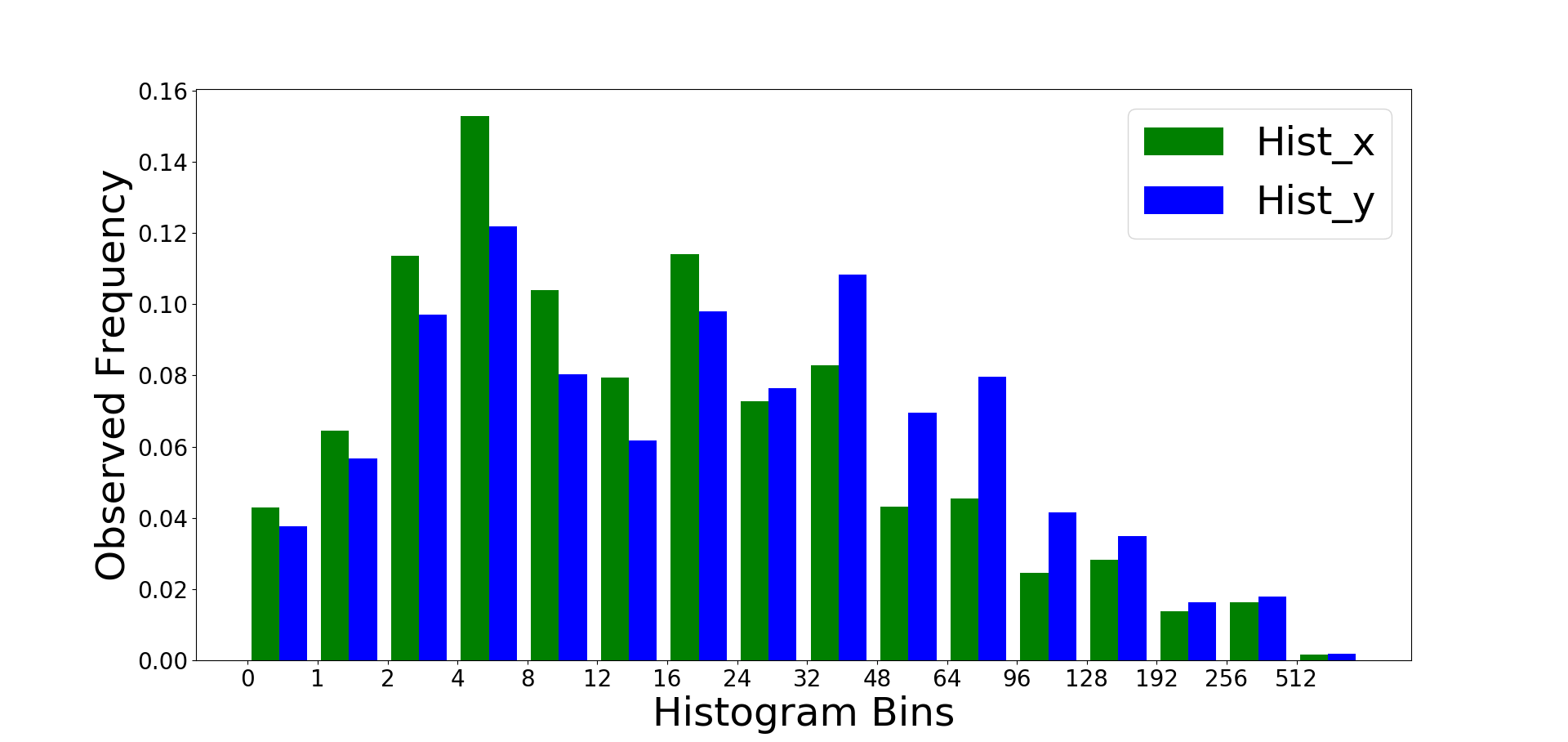}
}
\subfloat[Histograms in each Patch]{
\includegraphics[width=.4\textwidth]{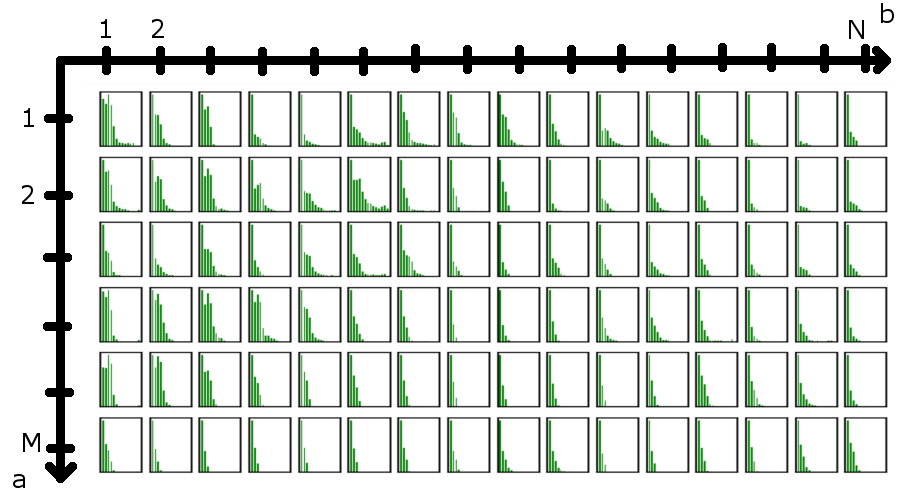}
}
\caption{Feature Representation for Image \ref{fig:patches}}
\label{fig:hist}
}
\end{figure}

The gradient value is set as stimulus intensity and categorized according to the Fechner law \cite{hecht1924fechner}. 
The Fechner law postulates that the actual and perceived change of stimulus intensity scales logarithmically.
The domain of [0,1020] is split into bins using powers of 2 as thresholds. 
Each bin edge represents approximately a linear increase in the perceived strength of a gradient stimulus. 
5 intermediate thresholds are added in the middle of the domain to increase resolution to a total of 16 bins.
In this fashion all gradient distributions $P_{ab}(g_x)$ are transformed into histograms $Hist_{ab}(g_x)$ (likewise $Hist_{ab}(g_y)$).
The result of such a binning is displayed in figure \ref{fig:hist}.
For each patch in the grid the histogram of gradients is collected.
It gives an indication of the amount of strong edges, smooth surfaces, shadows and visible content in that image patch.

\subsection{Interaction between Gradients}
Until now the marginal distributions $P_{ab}(g_x)$ and $P_{ab}(g_y)$ were viewed independently.
A dependency between $g_x$ and $g_y$ is however expected since they are partly computed on the same pixels.
Some IQA methods therefore utilize the gradient magnitude and orientation to take the information of the joint distribution  $P_{ab}(g_x,g_y)$ into account \cite{zhang2011fsim,chen2006gradient_ssim}.

To assess the gains of transmitting the joint distribution we examined the dependency of $g_x$ and $g_y$ in three IQA databases.
Table \ref{tab:gradient correlation} shows the median Pearson and Spearman correlation coefficients observed for the gradients $g_x$ and $g_y$.
The correlations deviate around zero.
This implies that there likely is no linear or monotone relationship between $g_x$ and $g_y$.
Due to the discrete nature of the histograms, mutual information is another tool to assess the information contained in the joint distribution $P(g_x,g_y)$. 
The Information Quality Ratio (IQR \cite{wijaya2017information}) of $0.3$  implies that there is relatively little information in the joint histogram compared to the marginals 
(MI: Mutual Information, H: Entropy).
\begin{equation}
IQR(g_x,g_y) =  \frac{MI(Hist(g_x),Hist(g_y))}{H(Hist(g_x),Hist(g_y)}
\end{equation}

Sending over a joint histogram $Hist_{ab}(g_x,g_y)$ with $16^2=256$ bins would cause a lot of extra overhead. 
However it contains only a modest amount of mutual information (30\% of marginals information).
Therefore we restrict ourselves to transmitting the two marginal distribution histograms  $Hist_{ab}(g_x)$ and $_{ab}Hist(g_y)$ for together $2*16 = 32$ bins.

\begin{table}[t]{
\caption{Correlation between $g_x$ and $g_y$}
\label{tab:gradient correlation}

\centering
\begin{tabular}{ | c | c | c | c | c |  }

\hline
  Dataset & Pearson Correlation & Spearman Correlation& IQR\cite{wijaya2017information} \\
 \hlineB{2.5}
 LIVE \cite{sheikh2005live} & $-0.0132(\pm 0.0773)$ &  $-0.01063(\pm 0.0651)$ & $0.2713(\pm 0.0208)$  \\
 \hline
 CSIQ \cite{larson2010csiq} & $-0.0212(\pm 0.1115)$ & $-0.0077(\pm 0.0714)$ & $0.2799(\pm 0.0436)$  \\
 \hline
 TID2013 \cite{ponomarenko2015tid} &  $-0.0192 (\pm 0.0777)$ & $-0.0207(\pm 0.0792)$ & $0.2756(\pm 0.0163)$  \\
 \hline
 
\end{tabular}

}

\end{table}

The complete feature set for an image consists of M x N patches containing two 16-bin histograms $Hist_{ab}(g_x)$ and $Hist_{ab}(g_y)$ of the Sobel-X/Y gradient distributions.
For lossless histogram transmission of an image with \textit{s} pixels the bins must have $\lceil log_2(s)\rceil$ bits.
In an automotive HD-image (1920x720) divided into 6x16 patches our features use $6*16*32*14 \text{Bit} \approx 5\textit{kB}$, compared to the full image with $1920*720*3*8 \text{Bit} \approx 4\textit{MB}$. 
Such feature data can easily be transmitted alongside a pixel stream. 
The histogram itself is a compact image signature that acts as a logarithmically contrast response model for gradients. 

\section{Distances for Contrast Distributions}
In the previous section we have looked extensively at the design of the histogram features.
This section aims to use the histogram grid for a perceptual comparison of images in regards to safety hazards.
Given the features of the \textit{reference} image $Histr_{ab}(g_x)$ and $Histr_{ab}(g_y)$ and the features of a \textit{processed} image $Histp_{ab}(g_x)$ and $Histp_{ab}(g_y)$, the goal is to detect and evaluate safety hazards resulting in a deterioration of quality.
The distance functions presented in the following are executed once each on the $g_x$ and $g_y$ histograms.

\subsection{Perceptual Dissimilarity}
\begin{figure}[t]
\centering
\subfloat[Distorted Image]{
\centering
\includegraphics[width=0.7\linewidth]{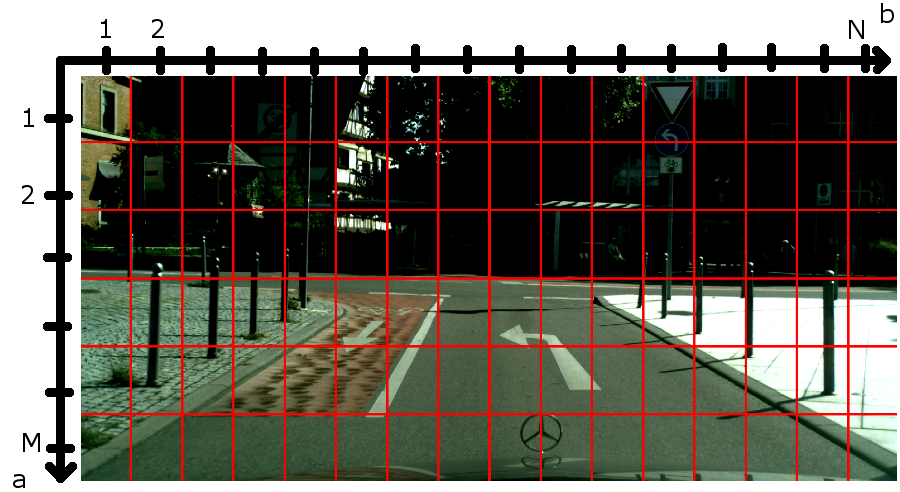}}

\subfloat[Histogram Comparison]{
\centering
\includegraphics[width=.5\linewidth]{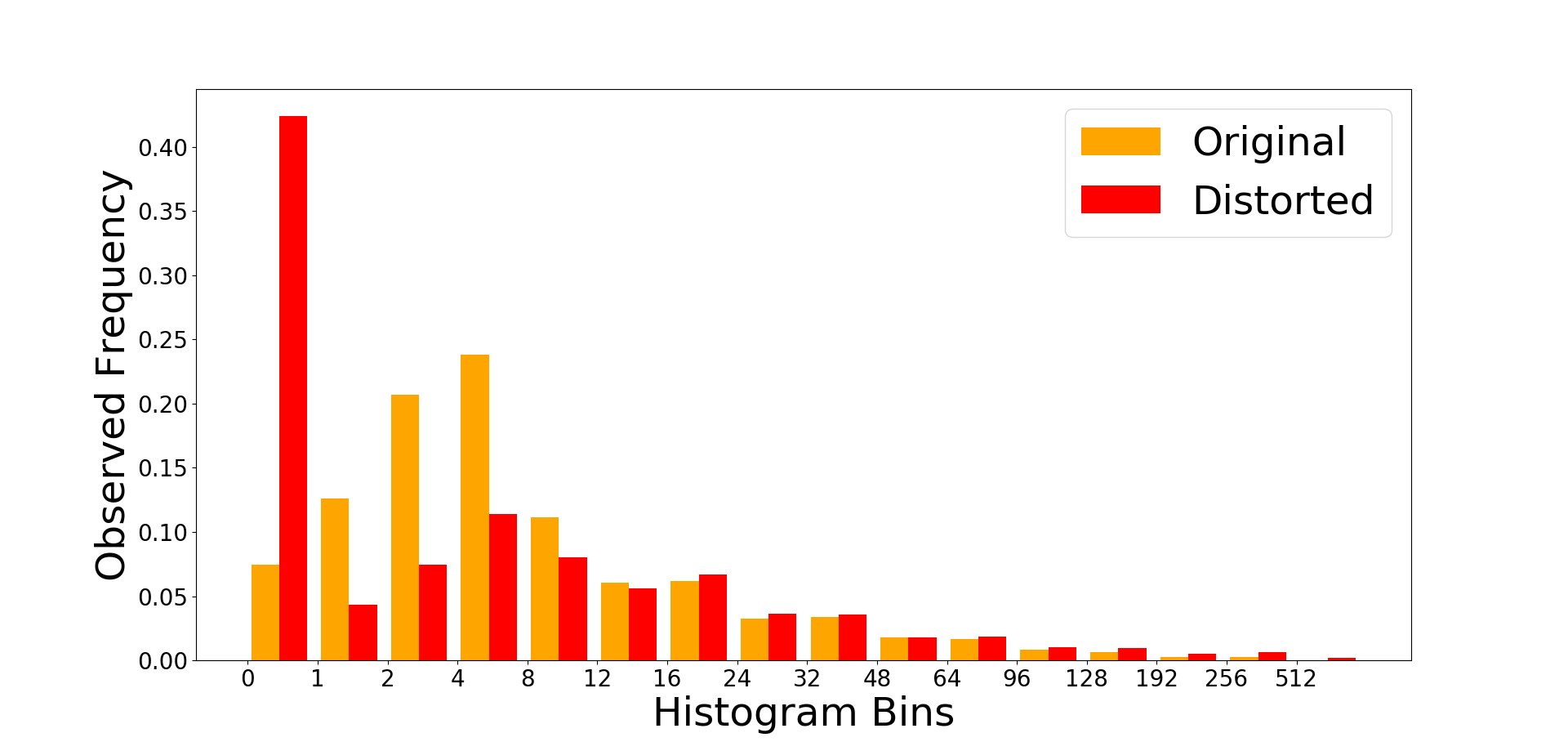}
\label{fig:hist_comp_bad}}
\subfloat[KL-divergence per Patch]{
\centering
\includegraphics[width=.5\linewidth]{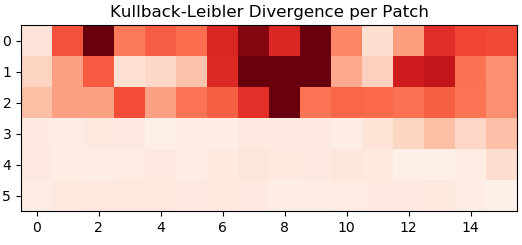}
\label{fig:heatmap}}

\caption{Effects of Image Distortion on $Hist(g_x)$}
\label{fig:hazards}
\end{figure}

When objects or pedestrians vanish from perception due to a faulty processing operation, a safety hazard has occurred and is to be detected.
Figure \ref{fig:hazards} shows an overly contrast stretched version of the image from figure \ref{fig:patches}.
In particular in the upper part of the image details are clipped away due to the processing.
A look at the computed distance map in figure \ref{fig:heatmap} highlights the advantages of our approach.
It clearly identifies the areas where most of the content was lost and the areas which are still similar to the reference.
Although the pixel values in the lower image half differ a lot from the reference, which would alarm PSNR for example, the gradient histograms are similar.
Thus our algorithm can infer that the lower image half is still perceptually similar.

\begin{table}[hb!]
    \caption{Histogram Distances ($Hist(i)$: value of i-th histogram bin)}
    \label{tab:dist_summary}
    \centering
    \begin{tabular}{l c}
    \\
    \hlineB{2.5}
    Kullback-Leibler$_{ab}$:$\sum_{i=1}^{16} Histr_{ab}(i) * log(\frac{Histr_{ab}(i)}{Histp_{ab}(i)})$\\
    \hline
    \\Emd$_{ab}$: $\sum_{i=1}^{16}| Emd_i |, \text{Emd}_{i+1} = Histr_{ab}(i) + Emd_{i} - Histp_{ab}(i)$ \\
    \hline
    \\Intersection$_{ab}$: $\sum_{i=1}^{16} min(Histr_{ab}(i),Histp_{ab}(i))$ \\
    \hline
    \\TotalVariation$_{ab}$: $max( | Histr_{ab}(i) - Histp_{ab}(i) | )$\\
    
    \hlineB{2.5}

\end{tabular}
\end{table}

Any significant change in the perception of the scene will result in an altered gradient distribution, may it be a vanishing of objects, the injection of noise or compression related artifacts.
To quantify these distortions, distance functions of probabilistic background are picked and designed.

In diagram \ref{fig:heatmap} the Kullback-Leibler divergence is used.
It provides an information-theoretic quantification of the difference between two histograms $Histr_{ab}$ and $Histp_{ab}$ of an image patch.
A successful contrast enhancement would sharpen only existing boundaries and clip away no details.
Since KL is weighted by the probability in each bin and edges are sparse, the resulting KL-divergence is negligible.
Another useful indicator is the Earth-Movers-Distance (Emd). 
It describes the amount of contrast that needs to be shifted to transform the histogram of the reference image to the result. 
The set of distances is rounded out with the Histogram Intersection and the Total-Variation distance, which simply describes the maximum bin difference between $Histr_{ab}$ and $Histp_{ab}$.
Together these distances give different estimates of perceptual similarity or degradation.

\begin{figure}[t]
    \centering
\subfloat[Noisy Image]{
\centering
\includegraphics[width=0.4\linewidth]{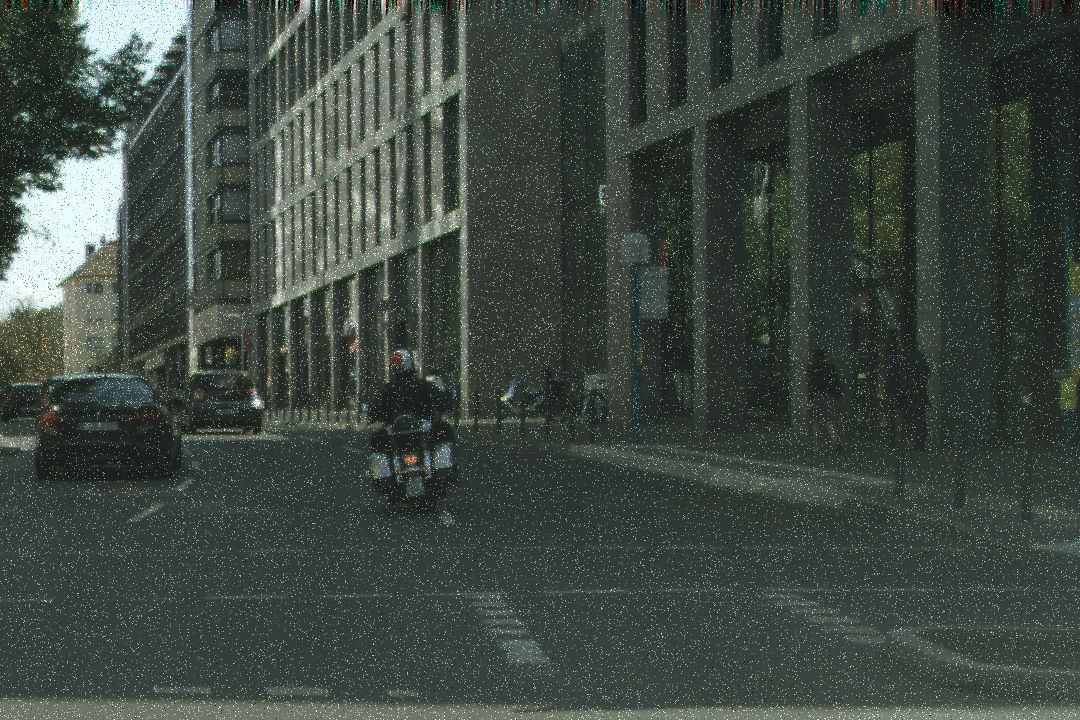}}
\subfloat[Histogram Comparison]{
\centering
\includegraphics[width=.6\linewidth]{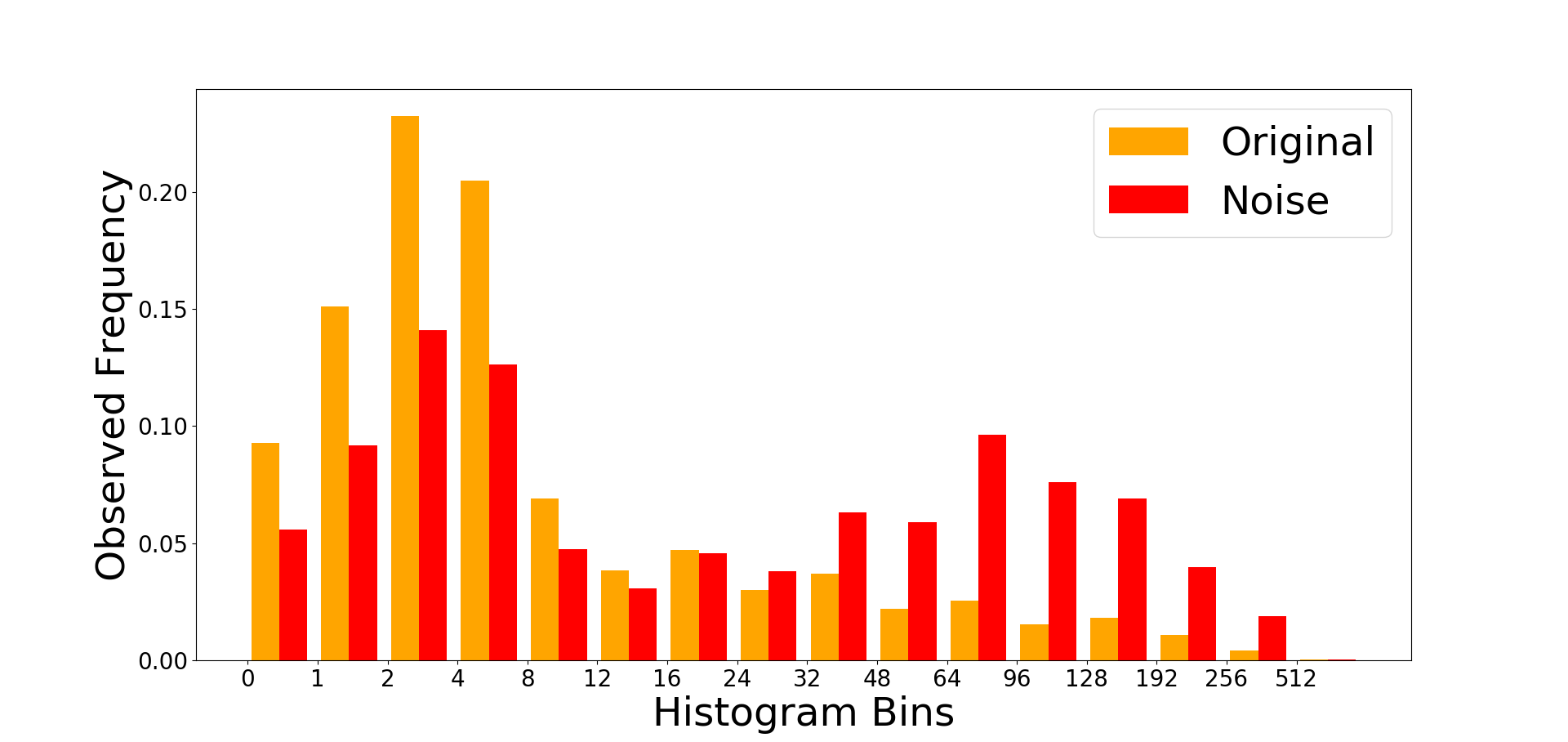}}
    \caption{Effects of White Noise on $Hist(g_x)$}
    \label{fig:noise_example}
\end{figure}

\subsection{Noise Artifacts}
The distances from table \ref{tab:dist_summary} treat each bin equally.
However some distortions mainly affect the specific lower/higher ranges of the histogram, which motivates to design custom distance functions.
When noise appears on a previously homogeneous surface, it produces strong, non-natural edges and gradients.
Figure \ref{fig:noise_example} shows the change after noise is introduced. 
There is a significant increase in the high contrast proportion of the histogram.

Due to its commonness, a custom distance function targeted at noise is added.
Insertion of noise is gauged by thresholding and comparing the amount of large gradients. 
The upper t bins of each histogram are summed up.
Two t values (4 and 6) are applied for the sake of robustness.
\begin{equation}
NoiseInc = \sum_{i=16-t}^{16} Hist_{ref}(i) - \sum_{i=16-t}^{16} Hist_{pro}(i)
\label{dist:high}
\end{equation}

\begin{figure}[t]
    \centering
\subfloat[JPEG Compressed]{
\centering
\includegraphics[width=0.4\linewidth]{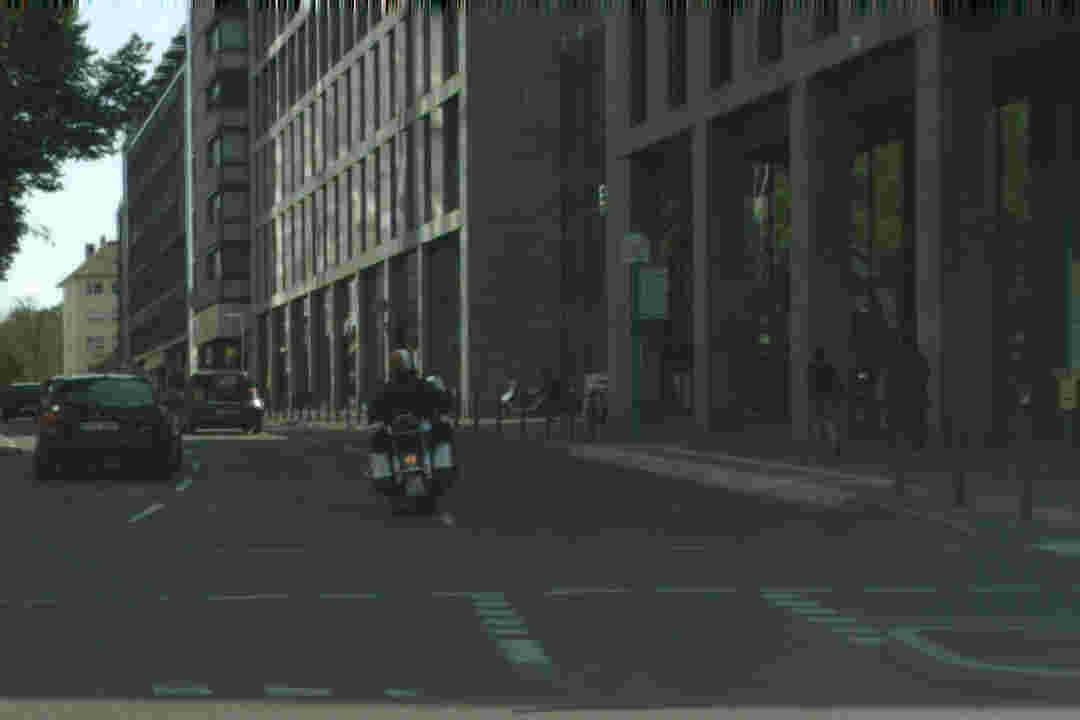}}
\subfloat[Histogram Comparison]{
\centering
\includegraphics[width=.6\linewidth]{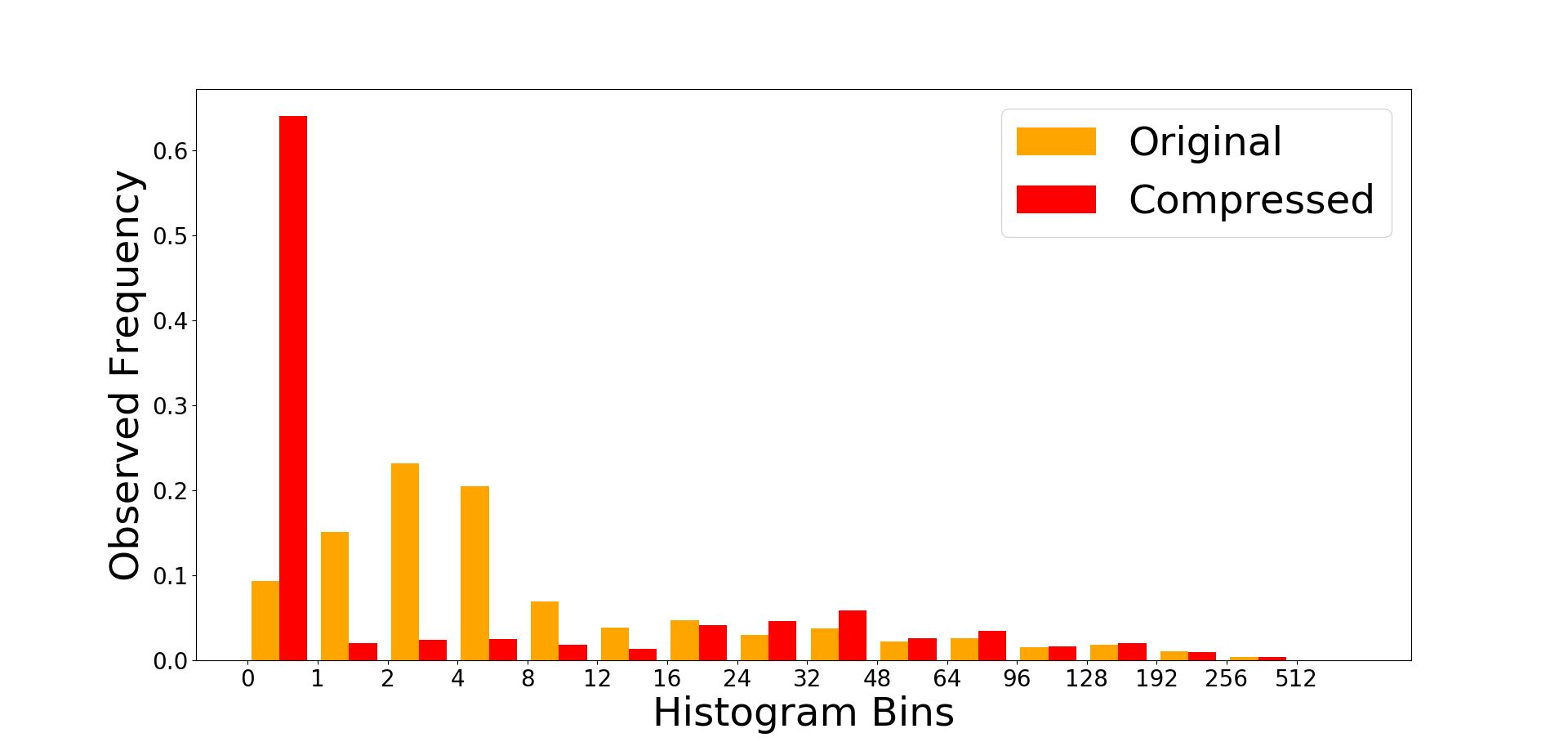}}
    \caption{Effects of JPEG Compression on $Hist(g_x)$}
    \label{fig:jpeg_hist}
\end{figure}

\subsection{Compression Artifacts}
Compression, specifically JPEG compression also has a special effect on our gradient histograms.
The JPEG \cite{wallace1992jpeg} standard uses the discrete cosine transform (DCT) to encode 8x8 pixel blocks. 
DCT is widespread and also used by video coding schemes like MPEG.
JPEG compression causes blocking: Small DCT coefficients are quantized to zero resulting in many blocks of same pixel values and zero gradients. 
Figure \ref{fig:jpeg_hist} shows the histogram shift caused by JPEG compression.  
The amount of blocking is strongly reflected in an increased proportion of zero gradients. 
They are contained in the first histogram bin.
\begin{equation}
Blocking = Histr_{ab}(1) - Histp_{ab}(1)
\label{dist:block}
\end{equation}

Furthermore it is also visible that the gradients are very unevenly distributed in the compressed image. 
The variety and range of contrast can be represented by the Shannon entropy of a histogram.
By comparing the respective entropy before and after processing, a shrink in the range of contrasts is detected which is often associated with visual distortion and artifacts.
\begin{equation}
    EntropyGap  = \sum_{i=1}^{16} Histr_{ab}(i) - \sum_{i=1}^{16} Histp_{ab}(i) 
    \label{dist:ent}
\end{equation}

All presented distance functions capture different aspects important for distortions such as loss of detail or noise.
Two variants of combining these distances into a system which predicts the strength of error in an image are explored:
\begin{enumerate}
    \item \textit{Spatial combination}: The distances obtained by computing the KL-divergence on each patch separately are aggregated into a global distortion score forming the method \textbf{CD$^2$-A}.
    \begin{equation}
        \sum_{a=1}^M \sum_{b=1}^N Kl_{ab}(g_x) + Kl_{ab}(g_y) 
    \end{equation}
    \item \textit{Functional combination}: 
    The set of all presented distance functions $d_i$ is to be transformed by a function \textit{f} into a quality score.
    The distances are computed only once per image on the global histograms for $g_x$ and $g_y$.
    \textit{f} is modelled through gradient tree boosting from the Lightgbm framework \cite{ke2017lightgbm}.
    This model (called \textbf{CD$^2$-B}) is trained with the help of IQA datasets and evaluated in the next section.
    An approach involving both a spatial and functional combination of distances is intriguing but left to future work.
    \begin{equation}
         f(d_1,...,d_{16}) = \sum_{i} \gamma_i * tree_i(subset_i(d_1,...,d_{16}))
    \end{equation}
\end{enumerate} 

\section{Experiments}
The proposed methods are tested on the LIVE \cite{sheikh2005live} , CSIQ \cite{larson2010csiq} and TID2013 \cite{ponomarenko2015tid} datasets. 
These datasets contain pairs of a reference and a processed image subject to different levels of compression, noise, etc. 
Human subjects give opinions about image quality that are transformed to difference-in-mean-opinion-score (DMOS) to quantify distortion.

CD$^2$-A/B is compared against the FR-IQA metrics PSNR, SSIM \cite{wang2004ssim} and FSIM \cite{zhang2011fsim} as well as the  RR-IQA indices RRED \cite{soundararajan2011rred} and WNISM \cite{wang2005reducedwavstat}. 
CD$^2$ is implemented in Python, for the other methods their original Matlab source code was used.
The DMOS prediction is tested with 5-fold cross validation.
Reported are the mean root mean-square error (RMSE), Pearson linear correlation coefficient (PLCC) and Spearman rank-order correlation coefficient (SROCC).

\begin{table}[t]
\caption{Performance on IQA-Datasets}
\label{tab:performance}
\centering
\begin{tabular}{| c c c | c | c | c || c | c | c | c | c | }
 \hline
    & & CD$^2$-A & CD$^2$-B & WNISM & RRED & PSNR & SSIM & FSIM \\
 \hlineB{2.5}
 \multirow{3}{*}{TID2013 \cite{ponomarenko2015tid}} 
 & RMSE & 0.9804 & \textbf{0.5659} & 1.0042 & 0.7134 & 1.0890 & 0.9027 & 0.5967  \\
 & PLCC & 0.6125 & \textbf{0.8962} & 0.5851 & 0.8180 & 0.4784 & 0.6868 & 0.8767 \\
 & SROCC & 0.5342 & \textbf{0.8891} & 0.5155 & 0.7633  & 0.6391 & 0.6264 & 0.8511 \\
 \hlineB{2.5}
 \multirow{3}{*}{LIVE \cite{sheikh2005live}} 
 & RMSE & 14.6130 & 7.8534 & 18.4173 & 9.4880 & 13.3790 & 11.6272 & \textbf{7.5718}  \\
 & PLCC & 0.8455 & 0.9595 & 0.7371 & 0.9383 & 0.8715 & 0.9056 & \textbf{0.9612} \\
 & SROCC & 0.8546 & 0.9582 & 0.7407 & 0.9397 & 0.8728 & 0.9098 & \textbf{0.9608} \\
 \hlineB{2.5}
 \multirow{3}{*}{CSIQ \cite{larson2010csiq}} 
 & RMSE & 0.1673 & \textbf{0.0925}  & 0.1934 & 0.1223 & 0.1736 & 0.1510 & 0.1092 \\
 & PLCC & 0.7684 &\textbf{0.9376} & 0.6760 & 0.8837 & 0.7509 & 0.8176 & 0.9091 \\
 & SROCC & 0.7943 & \textbf{0.9339} & 0.6821 & 0.9073 & 0.8023 & 0.8353 & 0.9212 \\
 \hlineB{2.5}

\end{tabular}
\end{table}

\subsection{DMOS Prediction Performance}
Table \ref{tab:performance} shows the DMOS prediction performance across the benchmarks. 
The best method is highlighted in each row.
CD$^2$-A posts a competitive performance in comparison with other RR-IQA methods.
It outmatches WNISM, though a significant gap exists to RRED.
RRED however requires $n*\frac{32}{36}$ Bit of feature data for n pixels and drops down to a similar level when its feature data is reduced \cite{soundararajan2011rred}.
CD$^2$-A performs comparable to the FR-IQA baseline PSNR.
In addition the method brings along a number of advantages in computational and data efficiency and offers local interpretability (see figure \ref{fig:heatmap}).

The CD$^2$-B boosting model performs great across the board. 
It posts the best performance for TID2013 and CSIQ.
It surpasses its competitor RR-IQA methods and is on par with the FR-IQA state of the art FSIM.
This performance edge may be based in the ability to adapt to the specific distortions of each dataset through training on all distances.
While indices like SSIM or CD$^2$-A score identically for different datasets, CD$^2$-B adapts its scoring to the current benchmark and gains over its static competitors.

\subsection{User Test}
\begin{figure}[t]
    \centering
    \subfloat[10 Processing Levels of JPEG]{
    \includegraphics[width=0.45\linewidth]{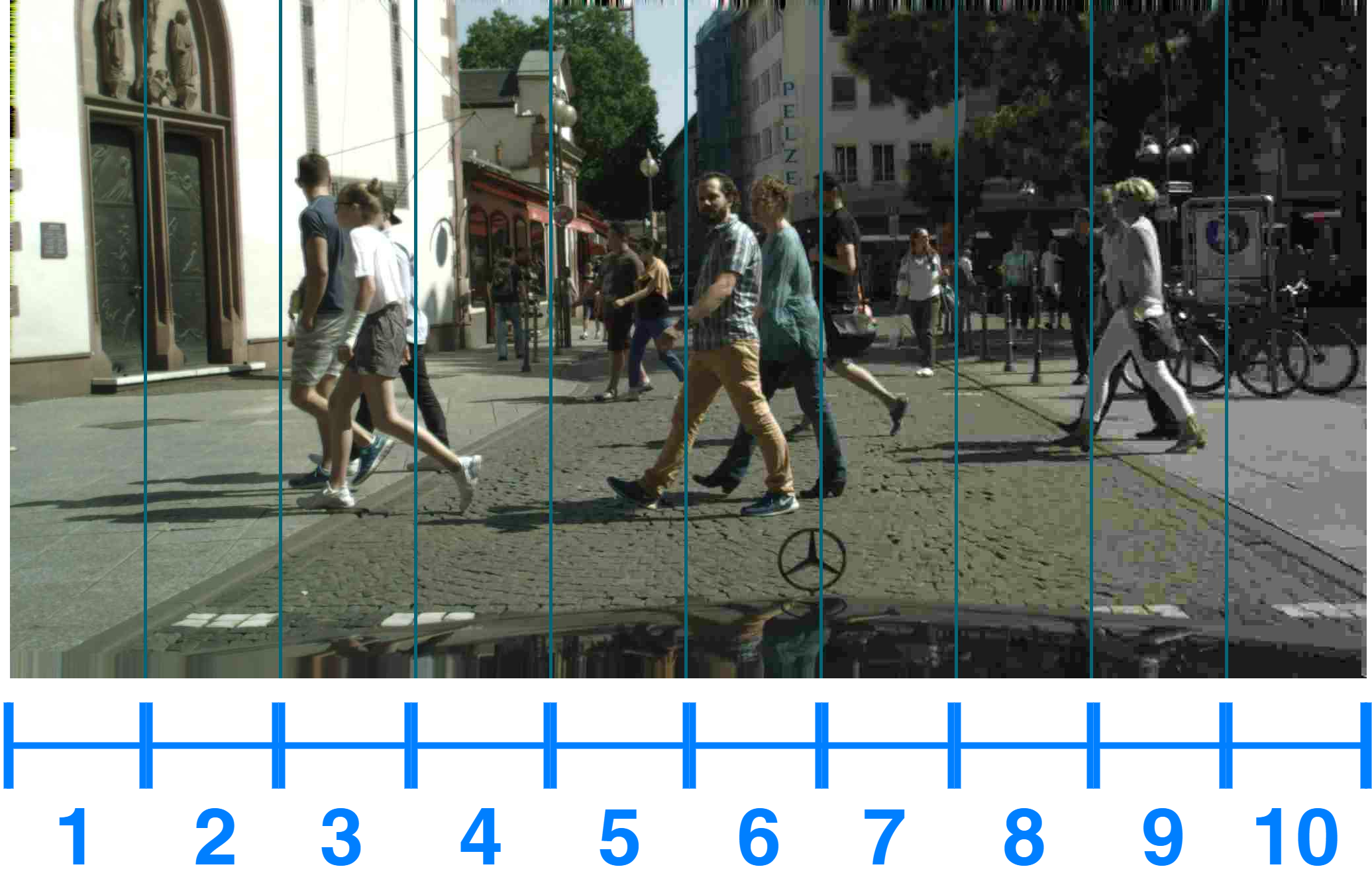}
    \label{fig:usertest_demo}
    }
    \subfloat[CD$^2$-A Scores of Unsafe Processing]{
    \includegraphics[width=0.55\linewidth]{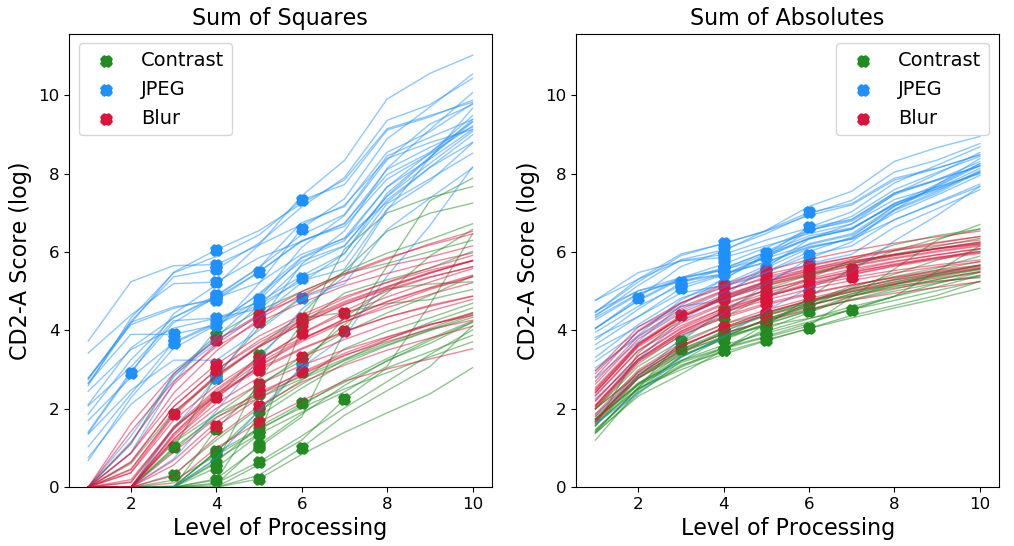}
    \label{fig:usertest_result}
    }
    \caption{Results of User Test}
    \label{fig:my_label}
    
\end{figure}

In addition an automotive application-oriented user test was carried out.
The goal of the test is to determine a threshold for CD$^2$-A at which an image distortion is deemed dangerous.
The subjects are set up in front of an automotive head-unit display producing 400 nits.
On the display street scenes from the Cityscapes dataset \cite{cordts2016cityscapes} are shown. 
The participants can adjust the strength of a processing operation (contrast, JPEG and gaussian blur) between ten levels, illustrated in figure \ref{fig:usertest_demo}. 
The task is to input the lowest level, where the subject is no longer able to perceive all important details of the scene.

In total 7 subjects (age 19-59, 2 female, 4 non-expert) participated in the study annotating 25 images per category.
The selected processing levels for each image were averaged to create pairs of original scene and unsafely processed scene.
Figure \ref{fig:usertest_result} shows the results of the test: Each point represents an image (X: average level, Y-value: CD$^2$-A score).
The lines visualize the CD$^2$-A score for all 10 levels.
Our first observation is that the level of dangerous processing varies based on content which justifies the approach of perceptual assessment instead.

The perfect perceptual score provides a threshold $\tau$ dividing safe from unsafe processing.
In the plot this would be a horizontal line $Y=\tau$ on which all points lie.
The two subplots of \ref{fig:usertest_result} compare taking the sum of squares of the KL-divergence grid (left) against the sum of absolutes (right).
The sum of absolutes provides a clearer vertical divide and was thus chosen for the CD$^2$-A method.
Per category the border between acceptable and dangerous levels of processing approximates a horizontal line, though the thresholds differ for different operations.
Therefore we see CD$^2$-A fit for use focused on assessing image processing if provided information about the type of the operation.

\section{Conclusion}
\label{chap:6}
We presented the CD$^2$ method for the perceptual assessment of image processing with regard to safety.
Based on local contrast histograms that categorize and compress contrast stimuli, a compact image signature is created.
At the image destination the quality of the altered image is verified with help of the original image signature.
Through a selection of statistical distances an accurate assessment about levels of noise, compression and detail loss can be made.
Tests on IQA benchmarks and a user study showcase the strong performance of our method.
While its feature data only uses 5kB, CD$^2$ is on par with the best FR-IQA methods that have access to millions of pixels.

At last we want to highlight the performance advantages when deploying $CD^2$ on integrated circuits.
The required logic for feature extraction and distance computation may be added to existing chips in cameras or displays for example.
This was simulated with a prototype on a Xilinx Artix-7 FPGA where less than 5\% of the logic gate resources were used.
Here the feature extraction of gradient histograms and the computation of the KL-Divergence take place on FPGA.
Since the the distribution of Sobel gradients can be aggregated on the fly, a three row pixel-buffer suffices to buffer an incoming video stream.
Feature extraction is finished $10 \mu s$ after the entire image is received and the feature distances may be computed in even less time. 
Deployed in modern automobiles, CD$^2$ is able to provide instant inference about image safety and hazards with modest means, even for 60fps HD-video.

%
%
\bibliographystyle{splncs04}
\bibliography{Bibliography}

\end{document}